\documentstyle[preprint,eqsecnum,aps,amsfonts]{revtex}
\newcommand{\R}{\Bbb R}       	
\newcommand{\e}{{\rm e}}
\begin{document}

\draft

\title          {The Dynamics of Multi-Scalar Field Cosmological Models and Assisted Inflation}

\author{        A A Coley }
\address{       Department of Mathematics and
                Statistics,
                Dalhousie University,
                Halifax, N.S., B3H 3J5}

\author{        R J  van den Hoogen}
\address{       Department of Mathematics, Statistics
                and Computer Science,
                Saint Francis Xavier University,
                Antigonish, N.S., B2G 2W5}

\maketitle

\begin{abstract}

We investigate the dynamical properties  of a class of spatially
homogeneous and isotropic cosmological models containing a
barotropic perfect fluid and multiple scalar fields with
independent exponential potentials. We show that the assisted
inflationary scaling solution is the global late-time attractor
for the parameter values for which the model is inflationary, even
when curvature and barotropic matter are included. For all other
parameter values the multi-field curvature scaling solution is the
global late-time attractor (in these asymptotically stable solutions   the
curvature is not dynamically negligible). Consequently, we find
that in general all of the scalar fields in multi-field models
with exponential potentials are non-negligible in late-time
behaviour, contrary to what is commonly believed. The early-time
and intermediate behaviour of the models is also studied. In
particular, $n$-scalar field models are investigated and the
structure of the saddle equilibrium points corresponding to
inflationary $m-$field scaling solutions and non-inflationary
$m-$field matter scaling solutions are also studied (where $m<n$),
leading to interesting transient dynamical behaviour with new
physical scenarios of potential importance.

\end{abstract}



\section{Introduction} \label{Introduction}

 Inflation is generally considered to be a reasonable solution to many of the
fundamental problems within the standard cosmological model.  In
original proposals \cite{Guth}, the early universe experiences a
period of accelerated expansion and essentially expands at an
exponential rate, (i.e., $R(t)\propto \e^{At}$ where $R(t)$ can be
considered as the size of the universe and $A$ is some positive
constant). Since these early stages, there have been a variety of
inflationary models that include scalar fields that have been
proposed \cite{Olive90}, and scalar fields have come to play an
important role in determining the dynamics of the early universe.
In one important class of inflationary models the condition of
exponential expansion is relaxed, and the universe grows at a
power-law rate, $R(t)\propto t^p$, where $p>1$ \cite{Luchin85}. In
particular, power-law inflationary models arise in models with a
scalar field $\phi$ having an exponential potential
$V(\phi)=V_0\e^{k\phi}$ \cite{Wetterich}. Although power-law
inflation is successful in solving the horizon and flatness
problems, inflation in these models persists into the indefinite
future and a phase
 transition
is required to bring
inflation to an end (however, see \cite{BC99}).

Spatially homogeneous  models
containing a scalar field $\phi$ with an exponential potential have been analyzed extensively
\cite{Halliwell87,Coley97a}.
It is known that all ever-expanding scalar field models  experience power-law
inflation when the parameter $k^2<2$; i.e., when the potential is sufficiently flat.
  The models have also been studied when $k^2>2$
\cite{Coley97a}.
Recently cosmological models containing both a scalar field with an
exponential potential and a perfect fluid with a
linear barotropic equation of state
have been studied.  It is found that in the general
class of Bianchi type B models that the power-law inflationary solution is
still the global attractor in the physically realistic regime (i.e., when
$\gamma>2/3$) if  $k^2<2$ \cite{Billyard99}.
Interestingly, the addition of a barotropic perfect
fluid creates the existence of a new type of solution, appropriately called a
matter scaling solution \cite{Copeland98,Billyard98,vandenHoogen99},
in which the energy-density of the scalar field scales with that of the matter; the effective
equation of state
for the scalar field is the same as that of the perfect fluid. The stability of
the matter scaling solution has been studied in \cite{Copeland98,Billyard98}.

Exponential potentials arise in many theories of the fundamental
interactions including superstring and higher-dimensional theories
\cite{Olive90,Green}. Typically, `realistic' supergravity theories
predict steep exponential potentials \cite{Green} (i.e., $k^2>2$),
effectively eliminating the possibility of power-law inflation.
However, dimensionally reduced higher-dimensional  theories also
predict numerous scalar fields, and so it is of interest to study
models with multiple scalar fields.

In the recent work of Liddle, Mazumdar and Schunck~\cite{Liddle98} the effect of additional scalar fields with
independent exponential potentials was considered. They assumed $n$ scalar fields
in a spatially flat Friedmann-Robertson-Walker (FRW) universe. They found that
an arbitrary number of scalar fields with
exponential potentials evolve towards a novel inflationary scaling solution,
which they termed
{\em assisted inflation}, in which all of the scalar fields
scale with one another (and are hence non-negligible asymptotically)
with the result that inflation occurs
even if each of the individual potentials is too steep to support
inflation on its own.
The existence
of multiple uncoupled scalar fields, each having an exponential potential, could therefore,
through a combined
(or {\em assisted}) effort, be a source for power-law inflation. This is true
even though
each individual scalar field need not be a source for inflation, and might therefore
lead to compatibility with supergravity theory.

In a recent dynamical analysis \cite{Malik98} it was shown that this assisted inflationary solution is a
late-time attractor in the class of zero-curvature FRW models. This was done by
choosing a redefinition of the fields (a rotation
in field space) which allows the effective potential
for field variations orthogonal to this solution to be written down; in analogy with models
of
hybrid inflation~\cite{hybrid1} it was then shown that
this potential has a global minimum along the attractor solution.
Also, analytic solutions describing homogeneous and inhomogeneous
perturbations about the attractor solution without resorting to slow-roll
approximations were presented in \cite{Malik98}, and curvature and isocurvature perturbation spectra
produced from vacuum fluctuations during assisted inflation were discussed.

In this paper we shall present a
qualitative analysis of models with the action

\begin{equation}
\label{action}
S=\int d^4 x \sqrt{-g} \left[ R
-\frac{1}{2} \sum_{i=1}^n (\nabla\phi_i^2)
-V_0 \sum_{i=1}^{n} e^{k_i \phi_i}\right] + S_m,
\end{equation}
where $S_m$ is the matter contribution.
Almost all previous analyses of multiple scalar field inflationary
models  have assumed zero-curvature
FRW spacetimes with no matter; here we extend the
analysis to include both curvature and matter. In section \ref{model} we shall
present the governing equations for $n$ scalar fields with exponential potentials
and matter. In section \ref{twomodel} we shall study the two-scalar field model
 with no matter, and, in particular,
discuss the stability of the two-field assisted inflationary model. In section \ref{twomodelmatter} we shall
study the two-scalar field model with barotropic matter. In section \ref{threemodel} we shall discuss three- and
multi-scalar field models. In section \ref{conclusions} we present our conclusions.

\subsection{Generalized Assisted Inflation}

Recently, models  with $n \times m$ scalar fields $\phi_{ij}$ and containing multiplicative
exponential terms in the effective potential of the form
\[
V_{eff}\equiv\sum_{i=1}^n \prod_{j=1}^{m}V_0\e^{k_{ij}\phi_{ij}}=\sum_{i=1}^n V_0^m\e^{\sum_{j=1}^m k_{ij}\phi_{ij}},
\]
where $1\leq i\leq n$ and $1\leq j\leq m$ and $k_{ij}$ are $n \times m$
real positive constants which are not zero, have also been
studied. A qualitative analysis of the $m=1$ case has been given in \cite{vandenHoogen99d},
where  an analogy was made with the dynamics of
soft inflation \cite{Berkin90}.

In \cite{Copeland99} a class of spatially flat FRW
multi-scalar fields models with multiplicative exponential potentials were studied.
Potentials of this form are quite common in dimensionally reduced
supergravity models \cite{stelle,reall}.
Exact two-field and general $n$-field power-law scaling inflationary solutions were obtained,
which were demonstrated to be late-time attractors,  generalizing the
assisted inflationary solutions previously obtained \cite{Liddle98};
this behaviour was dubbed `generalized assisted inflation'.
It was shown that it is more difficult to obtain assisted inflation in
these generalized models with cross-couplings between the scalar fields
in the potential; the fields in any one exponential term tend to conspire to act
against one another rather than assist each other (a result also noticed in
\cite{reall}). However,
as with the original version of assisted inflation,
this inhibiting affect can be compensated for
if there are enough exponential terms present in the potential (i.e. if $n$ is
large enough) \cite{Liddle98,Copeland99}.

The dynamics of `generalized assisted inflation' was investigated in more detail in \cite{GL}.
By introducing field rotations, which results in the introduction
of two orthogonal fields one of which is massless and the
other posseses an exponential potential \cite{Malik98,GL},
the nature of the late--time
attractor solution in a particular class of models was determined.
A dimensionally reduced action resulting from within the context of
a generalised toroidal compactification of higher--dimensional fields
in Einstein gravity minimally coupled to massless scalar fields was shown
to give rise to a model of the form under investigation, and it was shown how
the addition of interactions between the fields impede
inflation in this model.

Similar  behaviour was also noted by Kanti and Olive \cite{ko}
in multi-field assisted inflationary models with standard chaotic polynomial (rather than exponential)
potentials, which can arise in modern Kaluza-Klein
theories (and are a natural outcome of the compactification of
higher dimensional theories down to four dimensions). Indeed, Kanti and
Olive \cite{Kanti99} have recently proposed a possible realisation of assisted inflation based
on the compactification of a five-dimensional Kaluza Klein model, and have shown how
the additional fields of the assisted sector actually impede
inflation (they also showed that the assisted sector, coming
from a Kaluza Klein compactification,  eliminates the need for a fine-tuned quartic
coupling to drive chaotic inflation).  In Kaloper and Liddle \cite{KL} the dynamics of a simple implementation of the idea in Kanti and Olive \cite{Kanti99} was analyzed in more detail.  Since assisted inflation no longer corresponds to an asymptotic attractor, they found that as inflation proceeds the number of fields participating in the assisted behaviour decreases resulting in the interesting novel feature that the density perturbations generated retain some information about the initial conditions.


\section{The Model} \label{model}

We shall assume that the spacetime is spatially homogeneous and isotropic.  The line element for such a spacetime has the form
\[ds^2=-dt^2+R^2(t)\left[\frac{dr^2}{1-kr^2}+r^2(d\theta^2+\sin^2(\theta)d\phi^2)\right]
\]
where $k=+1,-1,0$ determines whether the model is  closed (positive-curvature), open (negative-curvature), or flat (zero-curvature).

We shall consider $n$ scalar fields $\phi_{i}$, where $1\leq i\leq n$, in which the effective potential has the form
\[
V_{eff}\equiv\sum_{i=1}^n V_0\e^{k_{i}\phi_{i}},
\]
where the $k_{i}$ are real non-zero positive constants.  We also assume that there exists a non-interacting perfect fluid
with density $\rho$ and pressure
\begin{equation}
p=(\gamma-1)\rho,\label{es}
\end{equation}
and we shall assume that $1\leq\gamma<2$.
The Einstein field equations, the conservation equations, together with the Klein-Gordon equations for the scalar fields,
 yield the following autonomous system of ordinary differential equations:
\begin{eqnarray}
H^2-\frac{1}{6}\left(\sum_{i=1}^{n}\dot\phi_{i}^2\right)-\frac{1}{3}V_{eff}-\frac{1}{3}\rho=-\frac{^3\!R}{6}, \nonumber\\
\dot{H}=-H^2 - \frac{1}{3}\left(\sum_{i=1}^{n}\dot\phi_{i}^2\right)+\frac{1}{3} V_{eff} -\frac{1}{6}(3\gamma-2)\rho,\label{system1}\\
\dot \rho=-3\gamma H \rho, \nonumber \\
\ddot\phi_{i}+3H\dot\phi_{i}+k_{i}V_0e^{k_{i}\phi_{i}}=0,\nonumber
\end{eqnarray}
 where $^3\!R=k/R^2$ is the curvature of the spacelike hypersurfaces, $H=\dot R/R$ is the Hubble expansion, and an overdot
 represents differentiation with respect to coordinate time $t$.  Units have been chosen so that $8\pi G = c =1$.

To analyze the dynamical system given in equation (\ref{system1}) we transform the system using expansion normalized variables.  Expansion normalized variables have had great success in analyzing the asymptotic behaviour of many cosmological models.  See \cite{Coley97a,WainwrightEllis97} for arguments in support of using dimensionless expansion normalized variables.  One primary reason involves the decoupling of one of the differential equations, which effectively reduces the dimension of the system by one, and, in some cases, compactifies the phase space.  We choose expansion normalized variables of the form
\begin{eqnarray}
\Omega=\frac{\rho}{3H^2},\qquad
\Phi_i=\frac{\sqrt{V_0}\e^{k_{i}\phi_{i}/2}}{\sqrt{3}H},\qquad
\Psi_{i}=\frac{\dot\phi_{i}}{\sqrt{6}H},\qquad \frac{dt}{d\tau}=\frac{1}{H}.\label{expansion}
\end{eqnarray}
  The resulting dynamical system describing these perfect fluid multiple scalar field models becomes
\begin{eqnarray}
\frac{d\Omega}{d\tau} &=& \Omega(2q-3\gamma+2),\nonumber \\
\frac{d\Psi_{i}}{d\tau} &=& \Psi_{i}(q-2)-\frac{\sqrt{6}}{2}k_{i}\Phi_{i}^2,\label{system}\\
\frac{d\Phi_i}{d\tau} &=& \Phi_i\left(q + 1 + \frac{\sqrt{6}}{2}k_{i}\Psi_{i} \right),\nonumber
\end{eqnarray}
for $(1\leq i \leq n)$, where the deceleration parameter has the following form
\[ q = \frac{(3\gamma-2)}{2}\Omega +2\sum_{i=1}^{n}\Psi_{i}^2-\sum_{i=1}^{n}\Phi_i^2,
\]
and
\[ \frac{^3R}{6 H^2}=-1 +\Omega +\sum_{i=1}^{n}\Psi_{i}^2+\sum_{i=1}^{n}\Phi_i^2.
\]

Assuming a non-negative energy density (i.e., $\Omega\geq0$) and
if $^3\!R \leq 0$, (i.e., in the negative and zero-curvature cases) the phase space for the dynamical system
in the expansion normalized variables $(\Omega,\Phi_i,\Psi_{i})$ is compact.  If $^3R>0$ (i.e., in the positive
curvature case) then the transformation given by equation (\ref{expansion}) becomes singular when $H=0$.  Here we
shall only make some partial comments with regards to the asymptotic behaviour of the positive curvature models.
All of the equilibrium points correspond to self-similar cosmological models and hence to power-law solutions
\cite{Billyard99}.


\section{Qualitative Analysis of Two-Scalar field Model} \label{twomodel}

 We shall first discuss the dynamics of the model with only two minimally coupled scalar fields
 and with no matter.
  We obtain this model by setting $n=2$ and $\Omega=0$ in (\ref{system}).
In this case we obtain the four-dimensional dynamical system
given by:
\begin{eqnarray}
\frac{d\Psi_{1}}{d \tau}&=&\Psi_{1}(q-2)-\frac{\sqrt{6}}{2}k_{1}{\Phi_1}^2 \nonumber\\
\frac{d\Psi_{2}}{d \tau}&=&\Psi_{2}(q-2)-\frac{\sqrt{6}}{2}k_{2}{\Phi_2}^2 \nonumber\\
\frac{d\Phi_{1}}{d \tau}&=&\Phi_{1}(q+1+\frac{\sqrt{6}}{2}k_1\Psi_{1}) \nonumber\\
\frac{d\Phi_{2}}{d \tau}&=&\Phi_{2}(q+1+\frac{\sqrt{6}}{2}k_2\Psi_{2})\label{DS2}
\end{eqnarray}
where
$$q=2\Psi_{1}^2+2\Psi_{2}^2-\Phi_{1}^2-\Phi_{2}^2$$
and
\[\frac{^3R}{6 H^2}=-1+\Psi_{1}^2+\Psi_{2}^2+\Phi_{1}^2+\Phi_{2}^2.\]

It is possible to choose simplified variables as in \cite{LuPope} via a rotation
in field space; although this would simplify the analysis of the assisted inflationary
solution, it would perhaps be more difficult to discribe all of the qualitative properties
of the models and relate this analysis to previous work.

\subsection{Assisted Inflation}

The flat {\em Assisted Inflation} model \cite{Liddle98} corresponds to
the equilibrium point $A$ of the system (\ref{DS2}) given by

\[ \{\Psi_1, \Psi_2, \Phi_1, \Phi_2\}^A \equiv \nonumber\]
\begin{equation}
       \{-\frac{k_1k_2^{\ 2}}{\sqrt{6}(k_1^{\ 2}+k_2^{\ 2})},
     -\frac{k_1^{\ 2}k_2}{\sqrt{6}(k_1^{\ 2}+k_2^{\ 2})},
     k_2\frac{\sqrt{6(k_1^{\ 2}+k_2^{\ 2})-k_1^{\ 2}k_2^{\ 2}}}
     {\sqrt{6}(k_1^{\ 2}+k_2^{\ 2})},
     k_1\frac{\sqrt{6(k_1^{\ 2}+k_2^{\ 2})-k_1^{\ 2}k_2^{\ 2}}}
     {\sqrt{6}(k_1^{\ 2}+k_2^{\ 2})} \}, \label{AAA}
\end{equation}
which is equivalent to
\[ \{-\frac{K^2}{\sqrt{6}k_1},-\frac{K^2}{\sqrt{6}k_2},\frac{\sqrt{K^2(6-K^2)}}{\sqrt{6}k_1},\frac{\sqrt{K^2(6-K^2)}}{\sqrt{6}k_2}\}\]
where
\[K^{-2}\equiv\frac{1}{{k_1}^{2}}+\frac{1}{{k_2}^{2}} .\]
The deceleration parameter for this solution is given by

\begin{eqnarray}
 q_A &\equiv& \frac{k_1^{\ 2}k_2^{\ 2}-2(k_1^{\ 2}+k_2^{\ 2})}{2(k_1^{\ 2}+k_2^{\ 2})}\nonumber \\
&=& \frac{K^2-2}{2}\label{qA}
\end{eqnarray}
and hence this solution, with
\[R(t) \propto t^p\nonumber\]
and
 \[ k_1 \phi_1 = k_2 \phi_2,\nonumber \]
is inflationary ($q_A<0$) if
\begin{equation}
 p \equiv 2 \sum_{i=1}^2 \frac{1}{{k_i}^2} = 2K^{-2}= \frac{1}{1+q_A} > 1; \qquad
 2 \geq K^{2}. \label{ppp}
\end{equation}

Since a single scalar field can only give rise to an inflationary power-law solution
if $\frac{1}{{k_i}^2}> \frac{1}{2}$ for $i=$ 1 or 2 \cite{Wetterich,Billyard99}, this means that the two-scalar field model can be inflationary
even when the each of the individual potentials is too steep for the corresponding single scalar field model to inflate
(and hence the terminology {\em assisted} inflation).
The eigenvalues corresponding to the equilibrium point $A$ are given by  (See Appendix for details.)
\begin{equation}
K^2-2, \quad \frac{K^2-6}{2}, \quad \frac{1}{4}\left((K^2-6)\pm\sqrt{(K^2-6)^2+8K^2(K^2-6)}\right) \label{eigen}
\end{equation}
Hence this equilibrium point is stable when (\ref{ppp}) is satisfied, and so the
corresponding assisted inflationary solution is a late-time attractor \cite{Malik98}.

\subsection{Stability of Equilibria}

We note that several of the equilibrium points occur in the three-dimensional invariant set
corresponding to
the zero-curvature models
defined by

\[1=\Psi_{1}^2+\Psi_{2}^2+\Phi_{1}^2+\Phi_{2}^2.\]
When matter is included, there exists a monotonic function so that in the full
dynamical phase space there can be no periodic or recurrent orbits
and the global dynamics can be determined.
This implies that the  qualitative
features  described in this
section can be more rigorously proven.
All of the equilibrium points and their corresponding eigenvalues are listed in Table \ref{table1}.
Using this  this table let us discuss the local stability of these equilibrium points.

As noted above the equilibrium point $A$, given by (\ref{AAA}), corresponds to the assisted
inflationary solution. It exists for all parameter values satisfying

\begin{equation} \frac{1}{6} < \frac{1}{{k_1}^{2}} + \frac{1}{{k_2}^{2}}, \label{e1} \end{equation}
and is a sink (late-time attractor) for all parameter values satisfying (\ref{ppp}) (else it is
a saddle).

There are two equilibrium points, denoted by $P_1$ and $P_2$, whose coordinate values and
associated eigenvalues are given in Table \ref{table1}, which correspond to zero-curvature
power-law solutions in which one scalar field (either $\phi_1$ or $\phi_2$, respectively)
is negligible; these solutions exist if $\frac{1}{6} < \frac{1}{k_i^2}$ and are inflationary if,
in addition, $\frac{1}{2} < \frac{1}{k_i^2}$ (for each $i=1,2$, respectively)
and correspond to the well-known single scalar field power-law solutions \cite{Luchin85,Wetterich}.
From Table \ref{table1} we see that each $P_i$ has two negative eigenvalues and one positive
eigenvalue for all relevent
parameter values and an additional eigenvalue which is negative if $k_i^2<2$ (and positive for
$2<k_i^2<6$); hence these points are saddles and have a one- or two-dimensional unstable manifold
depending upon whether $k_i^2<2$ or $k_i^2>2$, respectively.

There also exist equilibrium points, denoted by $CS_1$, $CS_2$ and $CS$,
 whose coordinate values and the
associated eigenvalues are given in Table \ref{table1}.
The solutions correspond to power-law solutions in which the curvature scales with the first scalar field, the second scalar field or both,
respectively. The single-field curvature scaling equilibrium points $CS_1$ and $CS_2$ are both saddles.
The two-field curvature scaling equilibrium point $CS$ is a sink whenever
$\frac{1}{2} > \frac{1}{k_1^2} + \frac{1}{k_2^2}$ (otherwise a saddle).  Whenever the two-field curvature scaling solution is stable, it necessarily has negative curvature.

There is an equilibrium point, denoted by $M$, corresponding to the Milne form of flat spacetime,
which is always a saddle.

Finally, there is a one-dimensional set of equilibrium points parametrized by $\Psi_0$, denoted by
$MSF$, corresponding to zero-curvature massless scalar field models (in which both potentials are
zero). There is one zero eigenvalue corresponding to the fact that there is a one-dimensional set of equilibrium points.
There are values for $\Psi_0$ for which the remaining three eigenvalues are positive and hence a subset
of $MSF$ are sources (the remainder are saddles). These correspond to well-known early-time attracting massless scalar
field  models \cite{Billyard99}.

\subsection{Discussion}

From the analysis above we conclude that the two-field assisted inflationary solution $A$ is the global attractor when
$\sum_{i=1}^2 k_i^{-2} > \frac{1}{2}$ and the two-field curvature scaling solution $CS$ is the global
attractor when
$\sum_{i=1}^2 k_i^{-2} < \frac{1}{2}$. The massless scalar field solutions $MSF$ are always the
early-time attractors.

In all cases both scalar fields are non-negligible in generic late-time behaviour. This is contrary to the
commonly held belief that in multi-field models with exponential potentials the scalar field
with the shallowest potential (i.e., smallest value of $k$) would dominate at late times. Indeed, we have shown that the single
field power-law inflationary models always correspond to saddles, so that we have the rather surprising
result that generically a single scalar field model {\em never} dominates at late-times!

We note that both the assisted inflationary solution and the massless scalar field early-time attractors
correspond to zero-curvature models. However, the curvature is not always dynamically negligible asymptotically
because the two-field curvature scaling solution has non-zero curvature.

There is a range of parameter values for which the assisted
inflationary solution is the global late-time attractor (when the
solution is non-inflationary it corresponds to a saddle). For all
of these parameter values the single field power-law solutions
$P_1$ and $P_2$ are saddles. However, there are allowable
parameter values for which either $P_1$ and $P_2$ are both
inflationary, or one is inflationary while the other is not, or
and both are non-inflationary. This might give rise to some new
interesting physical scenarios. For example, a model could
asymptote towards an inflationary single field solution $P_i$,
stay close to $P_i$ for an arbitrarily long period of time (since
$P_i$ is an equilibrium point) inflating all the time, and then
eventually leave $P_i$ and evolve towards the stable attracting
inflationary solution $A$. (Note that if either of $P_1$ or $P_2$
are inflationary, then $A$ is necessarily inflationary --see
\ref{ppp}). This is akin to a double-inflationary model
\cite{doubleinflate} in which the density fluctuations on large
and small scales decouple (i.e., the scale invariance of the
spectrum is broken) thereby allowing the possibility of more on
large scales which is in better accord with observations.


\section{Qualitative Analysis of Two-Scalar field Model with Matter} \label{twomodelmatter}

 To understand the underlying dynamics of the model with matter (i.e., with $\Omega\not=0$) we shall
 shall study the model with two minimally coupled scalar fields together with matter having
  energy density $\rho$ with the barotropic equation of state given by (\ref{es}).  This model
is obtained  by setting $n=2$ in (\ref{system}), whence we obtain the five-dimensional dynamical system  given by:
\begin{eqnarray}
\frac{d\Omega}{ d\tau} &=& \Omega(2q-3\gamma+2) \nonumber\\
\frac{d\Psi_{1}}{d \tau}&=&\Psi_{1}(q-2)-\frac{\sqrt{6}}{2}k_{1}{\Phi_1}^2 \nonumber\\
\frac{d\Psi_{2}}{d \tau}&=&\Psi_{2}(q-2)-\frac{\sqrt{6}}{2}k_{2}{\Phi_2}^2 \nonumber\\
\frac{d\Phi_{1}}{d \tau}&=&\Phi_{1}(q+1+\frac{\sqrt{6}}{2}k_1\Psi_{1}) \nonumber\\
\frac{d\Phi_{2}}{d \tau}&=&\Phi_{2}(q+1+\frac{\sqrt{6}}{2}k_2\Psi_{2})
\end{eqnarray}
where
$$q=\frac{3\gamma-2}{2}\Omega+2\Psi_{1}^2+2\Psi_{2}^2-\Phi_{1}^2-\Phi_{2}^2$$
and
\[\frac{^3R}{6 H^2}=-1+\Omega+\Psi_{1}^2+\Psi_{2}^2+\Phi_{1}^2+\Phi_{2}^2\]

\subsection{Invariant Sets and Monotonic Functions}
The zero-curvature models
constitute  a four-dimensional invariant set. The models with no matter also
constitute a four-dimensional invariant set.

The function
\begin{equation}
W=\frac{\Omega^2}{(\Omega+\Psi_{1}^2+\Psi_{2}^2+\Phi_{1}^2+\Phi_{2}^2-1)^2}
\end{equation}
has derivative
\begin{equation}
\frac{d W}{d\tau}=2(2-3\gamma)W.
\end{equation}
We observe that this function is monotonic when $\Omega\not = 0$  (i.e., non-zero matter)
and $(\Omega+\Psi_{1}^2+\Psi_{2}^2+\Phi_{1}^2+\Phi_{2}^2-1)\not=0$ (i.e., non-zero curvature).
We also observe that the sign of $3\gamma-2$ signifigantly changes the dynamics of these models.
For example, in the case of interest here  $3\gamma-2>0$, whence $W$ is a decreasing function of time $\tau$.
 This immediately implies that
\begin{itemize}
\item There exist no periodic or recurrent orbits in the full five-dimensional phase space
(this does not preclude the existence of closed orbits in the
invariant sets $\Omega=0$ and $^3R=0$;
however, we shall be primarily concerned with the dynamics of the models in the complete phase-space
with matter and non-zero curvature).
\item The future asymptotic state lies within the invariant set $\Omega=0$.
Matter becomes dynamically unimportant to the future.
\item The past asymptotic state lies within the set of zero-curvature models.
\end{itemize}

\subsection{Stability of Equilibria}

The equilibrium points can be classisfied into two sets; those with $\Omega=0$ and those with $\Omega\not = 0$.
All equilibrium points listed in Table \ref{table1} exist in this case with $\Omega=0$, and
Table \ref{table2} lists the equilibrium points with $\Omega=0$ together with the additional eigenvalue due to the addition of matter.
Using the function $W$ above,
we can further conclude that those equilibrium points in the
set $\Omega\not=0$ necessarily must have zero-curvature. Table \ref{table3}
lists the eigenvalues found in the invariant set $\Omega \not =0$.

Let us focus on the stability of the attractors in the full physical phase space. All late-time
attractors (sinks) occur in the invariant set $\Omega=0$. In the previous section we found that $A$ and $CS$
are the only sinks in the invariant set $\Omega=0$
(clearly, all of the saddles remain saddles in the full five-dimensional phase space). The additional eigenvalue
for the equilibrium point $A$ in the full physical phase space is given in Table \ref{table2} and is negative
if $\sum_{i=1}^2 k_i^{-2} > \frac{1}{3 \gamma}$. But this is always satisfied when
$\sum_{i=1}^2 k_i^{-2} > \frac{1}{2}$ and $\gamma > \frac{2}{3}$,
and hence $A$ is a sink and assisted inflation
is a global attractor.
Similarly, from Table \ref{table2} the equilibrium point $CS$ is always a sink for
$\sum_{i=1}^2 k_i^{-2} < \frac{1}{2}$ and hence the two-field curvature scaling solution remains the global
attractor in this case.

The early-time attractors lie in the zero-curvature invariant set and
consist of massless scalar field models. From Table \ref{table2} we see that
the massless scalar field models corresponding to the  repelling
equilibrium points $MSF$ are always sources (for $\gamma<2$).

\subsection{Matter Scaling Solutions}

In the case of a single scalar field
there exist zero-curvature FRW `matter scaling' solutions when the exponential potential is too steep to drive inflation,
in which the scalar field energy
density tracks that of the perfect fluid (so that at late
times neither field is neglible) \cite{Wetterich}.  In \cite{Copeland98}
it was shown that whenever these matter scaling
solutions exist they are the unique late-time attractors
within the class of flat
FRW models .  The cosmological consequences of these scaling
models have been further studied in \cite{FJ}.  For example, in
these models the scalar field energy density tracks that of the
perfect fluid and a significant fraction of the current energy
density of the Universe may be contained in the homogeneous scalar field
whose dynamical effects mimic cold dark
matter;  the tightest constraint on these cosmological models comes from
primordial nucleosynthesis bounds on any such relic density
\cite{Wetterich,Copeland98,FJ}.
The stability of these flat, isotropic matter scaling solutions was studied within
the class of spatially homogeneous cosmological models with a barotropic perfect
fluid and a scalar field with an exponential
potential in \cite{Billyard98}. It was found that
while the matter scaling solutions are stable to shear
perturbations, for
realistic matter with $\gamma \ge 1$ they are
unstable to curvature perturbations.

Returning to the models under investigation here, none of the equilibrium points with $\Omega\not=0$
can be late-time attractors for $\gamma > \frac{2}{3}$. Indeed, from Table \ref{table3} all such equilibrium points are seen to be
saddles. In particular, the two-field matter scaling solution corresponding to the equilibrium point
$MS$, which exists for
$\sum_{i=1}^2 k_i^{-2} < \frac{1}{3 \gamma}$, is a saddle. From Table \ref{table3}
we see that the first eigenvalue associated with $MS$ is positive, while the real parts of the remaining
four eigenvlaues are all negative. This is consistent with the stability analysis of
matter scaling solutions in models with a single scalar field  which found that the models were unstable to curvature
perturbations when $\gamma> \frac{2}{3}$ \cite{Billyard98}. However, these two-field
matter scaling solutions may still be of physical import. We note that when the curvature
is zero, the two-field matter scaling solution is an attractor (all four eigenvalues of $MS$ in the
four-dimensional zero-curvature invariant set have negative real parts -- so that $MS$ is a sink
in this invariant set), as in the case for the matter scaling solution in a single field model.
Note also from Table \ref{table3} that both of the single-field
matter scaling solutions, corresponding
to the equilibrium points $MS_1$ and $MS_2$, have two positive eigenvalues, so that again the solution
with multiple scalar fields is the `stronger' attractor.


\section{Qualitative Analysis of Three-Scalar Field Model} \label{threemodel}

Let us now consider models with more than two scalar fields. For simplicity, we
shall exclude a matter term here. However, from the previous section we easily determine
the essential properties resulting from the inclusion of a matter field. In particular, in this case
a monotonic function exists and this enables us to prove the qualitative results outlined below.
Let us begin with the three-scalar-field model,
obtained by setting $n=3$ and $\Omega=0$ in (\ref{system}).
In this case the resulting six-dimensional dynamical system is
given by:
\begin{eqnarray}
\frac{d\Psi_{1}}{d \tau}&=&\Psi_{1}(q-2)-\frac{\sqrt{6}}{2}k_{1}{\Phi_1}^2 \nonumber\\
\frac{d\Psi_{2}}{d \tau}&=&\Psi_{2}(q-2)-\frac{\sqrt{6}}{2}k_{2}{\Phi_2}^2 \nonumber\\
\frac{d\Psi_{3}}{d \tau}&=&\Psi_{3}(q-2)-\frac{\sqrt{6}}{2}k_{2}{\Phi_3}^2 \nonumber\\
\frac{d\Phi_{1}}{d \tau}&=&\Phi_{1}(q+1+\frac{\sqrt{6}}{2}k_1\Psi_{1}) \nonumber\\
\frac{d\Phi_{2}}{d \tau}&=&\Phi_{2}(q+1+\frac{\sqrt{6}}{2}k_2\Psi_{2})\nonumber\\
\frac{d\Phi_{3}}{d \tau}&=&\Phi_{3}(q+1+\frac{\sqrt{6}}{2}k_2\Psi_{3})\label{DS3}
\end{eqnarray}
where
$$q=2\Psi_{1}^2+2\Psi_{2}^2+2\Psi_{3}^2-\Phi_{1}^2-\Phi_{2}^2-\Phi_{3}^2$$
and
\[\frac{^3R}{6 H^2}=-1+\Psi_{1}^2+\Psi_{2}^2+\Psi_{3}^2+\Phi_{1}^2+\Phi_{2}^2+\Phi_{3}^2\]

Again it would possible to choose simplified variables as in \cite{LuPope} via a rotation
in field space as was done in recent work \cite{Malik98,GL}. However, we shall not do this
here. Indeed, we shall not present a complete qualitative analysis similar to that done in section \ref{twomodel},
since the essential features are similar and the detailed analysis would be long and painful.
Rather, let us describe the main effects of including a third scalar field on the inflationary solutions.

There exists a zero-curvature assisted inflationary solution which
now corresponds to the equilibrium point given by

\[   \{ \Psi_i=-\frac{K^{2}}{\sqrt{6}k_i},
 \Phi_i=\frac{\sqrt{K^{2}(6-K^{2})}}{\sqrt{6}k_i} \}  \]
where
\[K^{-2}\equiv k_1^{-2} + k_2^{-2} +k_3^{-2}. \]
In this solution all of the three scalar fields scale together at late times.
The corresponding eigenvalues are

\begin{eqnarray*}
K^{2}-2, \frac{K^{2}-6}{2},
\frac{1}{4}\left((K^{2}-6)\pm\sqrt{(K^{2}-6)^2+8K^{2}(K^{2}-6)}\right),\\
\frac{1}{4}\left((K^{2}-6)\pm\sqrt{(K^{2}-6)^2+8K^{2}(K^{2}-6)}\right).
\end{eqnarray*}
It is known \cite{Liddle98} to be a stable late-time attractor for all parameter values
for which the solution is inflationary (i.e., $K^{2}<2$; recall the
point does not exist if $K^{2}>6$).

There are three solutions in which two scalar fields scale together asymptotically and the third is negligible.
Assuming that the third scalar field is zero ($\Psi_{3}=\Phi_{3}=0$), the coordinates of the corresponding
equilibrium point, denoted by $P_{120}$, are given by

\begin{equation}
      \{-\frac{k_1k_2^{\ 2}}{\sqrt{6}(k_1^{\ 2}+k_2^{\ 2})},
     -\frac{k_1^{\ 2}k_2}{\sqrt{6}(k_1^{\ 2}+k_2^{\ 2})}, 0,
     k_2\frac{\sqrt{6(k_1^{\ 2}+k_2^{\ 2})-k_1^{\ 2}k_2^{\ 2}}}
     {\sqrt{6}(k_1^{\ 2}+k_2^{\ 2})},
     k_1\frac{\sqrt{6(k_1^{\ 2}+k_2^{\ 2})-k_1^{\ 2}k_2^{\ 2}}}
     {\sqrt{6}(k_1^{\ 2}+k_2^{\ 2})}, 0 \}. \label{BBB}
\end{equation}
Four of the eigenvalues are given by (\ref{eigen}), which all have negative real parts.

There are three solutions in which one scalar field scale dominates asymptotically and the remaining two are negligible.
Assuming that the first scalar field is non-zero ($\Psi_{1} \ne 0 \ne \Phi_{1}$), the coordinates of the corresponding
equilibrium point, denoted by $P_{100}$, are given by

\[ \{-\frac{k_1}{\sqrt{6}}, 0 , 0, \sqrt{1-\frac{k_1^{\ 2}}{6}}, 0, 0 \}. \]
Two of the eigenvalues are negative, one is positive, and there is an additional eigenvalue which
is negative if $k_1^2<2$ and positive if $2<k_1^2<6$.

In both of these cases the additional (remaining) two eigenvlaues can be calculated and are given by
\[ \{q-2<0, q+1>0 \}, \]
where $q$ is the deceleration parameter evaluated at the equilibrium point. Hence, the point $P_{120}$ is a saddle with one eigenvalue
with positive real part. The equilibrium points denoted by $P_{103}$ and $P_{023}$ are also saddles with one eigenvalue
with positive real part. In additional, the point $P_{100}$ is a saddle with two eigenvalues with positive real
parts (if $k_i^2<2$, and three eigenvalues with positive real parts if $k_i^2>2$). The same is true for the
equilibrium points denoted by $P_{020}$ and $P_{003}$.

Consequently there is a nested set of equilibrium points. At the top is the stable three-scalar field
assisted inflationary solution. In the next layer there are three two-scalar field models which are saddles
with  one eigenvalue
with positive real part. In the final layer there there are three one-scalar field models which are saddles
with  two eigenvalues
with positive real parts (or three eigenvalues
with positive real parts). Associated with this dynamical nesting are cosmological models with very
interesting physical properties.

This will follow through in the case of $n$ scalar fields. There will be a unique stable $n$-scalar field
assisted inflationary solution. There will then be $n$ of the ($n-1$)-scalar field models which are saddles
with  one eigenvalue
with positive real part. There will be $\frac{1}{2} n(n-1)$ of the ($n-2$)-scalar field models which are saddles
with  two eigenvalues
with positive real parts. And so on. Finally, there will be $n$ of the ($1$)-scalar field models which are saddles
with  $n-1$ (or $n-2$) eigenvalues
with positive real parts. As one `goes up' the nested structure the equilibrium points respectively become
`stronger attractors' (i.e., the stable manifold of the equilibrium points increases in dimension).

There is also a three-field curvature scaling solution corresponding to the
equilibrium point given by

\[ \{\Psi_i=-\frac{2}{\sqrt{6}k_i},\Phi_i=\frac{2}{\sqrt{3}k_i}\} \]
whose associated eigenvalues are given by

\[ -1\pm\sqrt{1+4K^{-2}(2-K^{2})} , -1\pm\sqrt{3}i , -1\pm\sqrt{3}i. \]
This equilibrium point is a
sink whenever $K^{2}>2$, in which case it represents an FRW model with negative
curvature $\frac{2-K^{2}}{K^{2}}$
 (else it is a saddle and represents a positive
curvature model).

Finally, there are saddle equilibrium points corresponding to the Milne model and the
one- and two-field
curvature scaling solutions, and a set of equilibrium points with
$\{\sum_{i=1}^{3}\Psi_i^{\ 2}=1,\Phi_i=0\}$ corresponding to massless scalar
field
models, a subset of which are sources.

A complete qualitative analysis can be done for $n$ scalar field models.
All of these results can be proven by induction (see, for example, \cite{Malik98}).
The $n$-scalar field
assisted inflationary solution is given by \cite{Liddle98}

\[R(t) \propto t^p\nonumber\]
and
 \[ k_i \phi_i = k_j \phi_j; \forall 1 \le i \ne j \le n,\nonumber \]
and

\[ p \equiv 2 \sum_{i=1}^n \frac{1}{{k_i}^2} > 1. \]
We note that in the two-scalar field model, although inflation can occur for potentials that are steeper
than in the single-field case, it cannot occur for arbitrarily steep potentials. For example,
if $k_1 = k_2 \equiv k$, then inflation occurs if $k^2<4$. However, for $n$-fields, if
$k_i = k$ for all $i$, then inflation occurs if $k^2<2n$; e.g., $k^2<8$ for four scalar field models.


\section{Conclusion} \label{conclusions}

We have studied multi-scalar field FRW cosmological models with
exponential potentials, extending previous analysis by including
non-zero curvature and barotropic matter. We have used dynamical
systems techniques, and by establishing a monotonic function in
the complete dynamical phase space (which includes both matter and
curvature), we have been able to deduce global results.

In section \ref{twomodel} a comprehensive qualitative analysis was presented in the case of two scalar fields
with no matter.
We concluded that the two-field assisted inflationary solution $A$ is the global attractor when
$\sum_{i=1}^2 k_i^{-2} > \frac{1}{2}$ and the two-field curvature scaling solution $CS$ is the global
attractor when
$\sum_{i=1}^2 k_i^{-2} < \frac{1}{2}$. A subset of the massless scalar field solutions $MSF$ are always the
early-time attractors.
Consequently, we found that in all cases both scalar fields are non-negligible in generic late-time behaviour; this
is an interesting and unexpected result and is contrary to the
commonly held belief that in multi-field models with exponential potentials the scalar field
with the shallowest potential would dominate at late times (indeed, we have shown that the single
field power-law inflationary models always correspond to saddles).
We note that both the assisted inflationary solution and the massless scalar field early-time attractors
correspond to zero-curvature models. However, the curvature is not always dynamically negligible asyptotically
because the two-field curvature scaling solution has non-zero curvature.

The zero-curvature assisted inflationary FRW scaling solutions \cite{Liddle98} are of particular importance
since, through the combined effect of multiple uncoupled scalar fields, each having an exponential potential,
power-law inflation is possible
even when
each individual scalar field need not be a source for inflation.
We have discussed the stability of the two-field assisted inflationary model, and generalized previous results by
including non-zero curvature to
show that for an appropriate
range of parameter values the assisted inflationary solution is the global late-time
attractor. For these parameter
values the single field power-law solutions $P_1$ and $P_2$ were shown to be saddles, and we showed that there are allowable parameter
values for which either $P_1$ and $P_2$ are both inflationary, or one is inflationary while the other is not, or
and both are non-inflationary, perhaps leading to
new interesting physical scenarios.

In section \ref{twomodelmatter} we
studied the two-scalar field model with barotropic matter. A monotonic function was established
in the resulting phase space. This proved that the matter must be negligible at late times and we
found that
$A$ and $CS$ are the only global sinks and that consequently assisted inflation and the two-field curvature scaling solution
are the global late-time attractors in their appropriate respective parameter ranges. This confirmed the result
that both scalar fields must be dynamically non-negligible in generic late-time behaviour, and
establishes the  stability of the two-field assisted inflationary model when matter is included.
The monotonic function also shows that
the early-time attractors lie in the zero-curvature invariant set, and we showed that they
consist of a subset of the massless scalar field models.

For $\gamma > \frac{2}{3}$, all of the equilibrium points with
$\Omega\not=0$ were shown to be saddles (see Table \ref{table3}).
The two-field matter scaling solution corresponding to the
equilibrium point $MS$ was shown to have a single positive
eigenvalue. Both of the single-field matter scaling solutions,
corresponding to the equilibrium points $MS_1$ and $MS_2$, were
shown to have two positive eigenvalues, so that again the solution
with multiple scalar fields is the `stronger' attractor. We note
that when the curvature is zero, the two-field matter scaling
solution is the late-time attractor, consistent with the stability
analysis in \cite{Billyard98}. These matter scaling solutions, and
particularly the two-field matter scaling solutions, give rise to
new transient dynamical behaviour and may be of physical import.
For example, there are solutions which spend a period of time with
the scalar field mimicking the barotropic fluid in which there is
a non-negligible scalar field (dark matter) energy density
(corresponding to a matter scaling saddle equilibrium point) and
subsequently evolve towards a scalar-field dominated power-law
inflationary epoch (corresponding to a single-field saddle
equilibrium point or a two-field assisted inflationary attractor)
with an accelerated expansion, perhaps explaining current high
redshift data.

 In section \ref{threemodel} we discussed three- and
multi-scalar field models (where, for simplicity, a matter term
was excluded). In the three-scalar field model we again
established the assisted inflationary solution and three-field
curvature scaling solutions as the stable late-time attractors. We
then considered  $n$-scalar field models, and established a nested
structure for the $m$-field scaling (assisted inflationary)
solutions. The $n$-scalar field assisted inflationary solution is
again the late-time attractor. All of the $m$-field (with $m<n$)
scaling solutions are saddles and in general the equilibrium
points corresponding to the $m$-field scaling solutions will have
$n-m$ eigenvalues with positive real parts so that the equilibrium
points corresponding to the greater number of non-negligible
scalar fields are, respectively,  the `stronger attractors'.
Again we should emphasize that Malik and Wands \cite{Malik98} showed that the multi-scalar field assisted inflationary solution is a late time attractor by utilizing a rotation in field space; indeed, the stable modes in a general stability analysis of this solution are presumably related to the isocurvature  perturbations orthogonal to the attractor trajectory in field space obtained in their analysis.

Finally, from previous investigations \cite{Billyard99} of
spatially homogeneous scalar field cosmological models with an
exponential potential and barotropic matter and from the above
analysis, we can conclude that the assisted inflationary solution
is a global attractor for all ever-expanding spatially homogeneous
multi-field cosmological models with exponential potentials
provided $\sum_{i=1}^{n}k_{i}^{-2}>1/2$.
 We can also conclude that the multi-field
curvature scaling solution is a global attractor for models of
Bianchi types $V$ and $VII_h$ provided
$\sum_{i=1}^{n}k_{i}^{-2}<1/2$ \cite{vandenHoogen97a} and a
multi-field generalization of the Feinstein-Ibanez anisotropic
single-field solution \cite{Feinstein93} is the global attractor for models of
Bianchi types $III$ and $VI_h$ if $\sum_{i=1}^{n}k_{i}^{-2}<1/2$.
Indeed, there will be $n$-field generalizations corresponding to
all equilibrium points of the single-field Bianchi Type B models
(cf. \cite{Billyard99}).

Finally, we note that spatially flat FRW matter scaling solutions
also exist in the context of generalized assisted inflation. In
\cite{GL} it was shown that in the higher-dimensional  context, in
the six--dimensional model the assisted dynamics between the
scalar fields mimics the behaviour of a relativistic fluid
$(\gamma =4/3$), while for higher dimensions the scalar fields
dominate the radiation component, perhaps leading to a `moduli'
problem for the early universe.


\acknowledgements

We would like to thank Nemanja Kaloper, Jim Lidsey and David Wands for interesting discussions, and
Laura Filion for help in checking the calculations.
AAC and
RJvdH are supported by grants from the Natural Sciences and Engineering Research Council of Canada.



\appendix
\section*{Determination of the Eigenvalues for the Equilibrium Point $A$}

If we let ${\bf X}=(\Psi_1,\Psi_2,\Phi_1,\Phi_2)$ then we are able to write the dynamical system (\ref{DS2}) as $d{\bf X}/d\tau={\bf F}({\bf X})$ where ${\bf F}$ is an analytic function from ${\R}^{4} \to {\R}^{4}$.  Standard results from dynamical systems theory states that the local behaviour near an equilibrium point, ${\bf X}_0$, of a system of non-linear autonomous differential equations of the form $d{\bf X}/d\tau={\bf F}({\bf X})$ is equivalent to the linearized system $d{\bf X}/d\tau=D{\bf F}({\bf X}_0){\bf X}$ in a neighborhood of the equilibrium point provided the eigenvalues of the matrix $D{\bf F}({\bf X}_0)$ have non-zero real parts \cite{DS}.  For the system given by equation (\ref{DS2}) the derivative matrix has the form
\begin{eqnarray*}
D{\bf F}({\bf X})= \left[\begin{array}{cccc}
  \! q-2+ 4\,{\Psi _{1}}^{2}  \, 							&
  \,4\,{\Psi _{1}}\,{\Psi _{2}}\,							&
  \, - 2\,{\Psi _{1}}\,{\Phi _{1}} - \sqrt{6}\,{k_{1}}\,{\Phi_{1}}\,			&
  \, - 2\,{\Psi _{1}}\,{\Phi _{2}} \!   						\\
  \! 4\,{\Psi _{1}}\,{\Psi _{2}}\,							&
  \, q-2+4\,{\Psi _{2}}^{2}  	\,							&
  \, - 2\,{\Psi _{2}}\,{\Phi _{1}}\,							&
  \, - 2\,{\Psi _{2}}\,{\Phi _{2}} - \sqrt{6}\,{k_{2}}\,{\Phi _{2}} \! 			\\
  \! {\Phi _{1}}\,(4\,{\Psi _{1}} + {\displaystyle \frac {\sqrt{6}}{2}}\,{k_{1}})\,	&     
  \,4\,{\Psi _{2}}\,{\Phi _{1}}\,							&
  \,q - 2\,{\Phi _{1}}^{2} + 1 + {\displaystyle \frac {\sqrt{6}}{2}} \,{k_{1}}\,{\Psi _{1}}\,											&
  - 2\,{\Phi _{1}}\,{\Phi _{2}} \! \!  							\\
  \! 4\,{\Psi _{1}}\,{\Phi _{2}}\,							&
  \,{\Phi _{2}}\,(4\,{\Psi _{2}} + {\displaystyle \frac {\sqrt{6}}{2}} \,{k_{2}})\,	&
  \, - 2\,{\Phi _{1}}\,{\Phi _{2}}\,&
  \, q-2\,{\Phi _{2}}^{2} + 1 + {\displaystyle \frac {\sqrt{6}}{2}} \,{k_{2}}\,{\Psi _{2}} \! \!  
\end{array} \right] 
\end{eqnarray*}
As an example, we shall evaluate $D{\bf F}({\bf X})$ at the assisted inflationary equilibrium point (\ref{AAA}) which is denoted as $A$.
\begin{eqnarray*}
D{\bf F}(A)=  \left[ \begin{array}{cccc}
{\displaystyle \frac {1}{2}} \,K^{2} + {\displaystyle \frac {2}{3
}} \,{\displaystyle \frac {K^{4}}{{k_{1}}^{2}}}  - 3 &
{\displaystyle \frac {2}{3}} \,{\displaystyle \frac {K^{4}}{{k_{1
}}\,{k_{2}}}}  & 
{\displaystyle \frac {1}{3}} \,{\displaystyle 
\frac {\bar K\,(K^{2} - 3\,{k_{1}}^{2})}{{k_{1}}^{2}}}  &
{\displaystyle \frac {1}{3}} \,{\displaystyle \frac {K^{2}\,
\bar K}{{k_{1}}\,{k_{2}}}}  \\ [2ex]
{\displaystyle \frac {2}{3}} \,{\displaystyle \frac {K^{4}}{{k_{1
}}\,{k_{2}}}}  & 
{\displaystyle \frac {1}{2}} \,K^{2} + 
{\displaystyle \frac {2}{3}} \,{\displaystyle \frac {K^{4}}{{k_{2
}}^{2}}}  - 3 &
{\displaystyle \frac {1}{3}} \,{\displaystyle 
\frac {K^{2}\,\bar K}{{k_{1}}\,{k_{2}}}}  &
{\displaystyle 
\frac {1}{3}} \,{\displaystyle \frac {\bar K\,(K^{2} - 3\,{
k_{2}}^{2})}{{k_{2}}^{2}}}  \\ [2ex]
{\displaystyle \frac {1}{6}} \,{\displaystyle \frac {\bar K
\,( - 4\,K^{2} + 3\,{k_{1}}^{2})}{{k_{1}}^{2}}}  &  
-{\displaystyle \frac {2}{3}} \,{\displaystyle \frac {K^{2}\,
\bar K }{{k_{1}}\,{k_{2}}}}  &
- {\displaystyle \frac {1}{3
}} \,{\displaystyle \frac {K^{2}\,(6 - K^{2})}{{k_{1}}^{2}}}  & 
 - {\displaystyle \frac {1}{3}} \,{\displaystyle \frac {K^{2}\,(6
 - K^{2})}{{k_{1}}\,{k_{2}}}}  \\ [2ex]
 - {\displaystyle \frac {2}{3}} \,{\displaystyle \frac {K^{2}\,
\bar K}{{k_{1}}\,{k_{2}}}}  &
{\displaystyle \frac {1}{6}} 
\,{\displaystyle \frac {\bar K \,( - 4\,K^{2} + 3\,{k_{2}}^{
2})}{{k_{2}}^{2}}}  &
  - {\displaystyle \frac {1}{3}} \,
{\displaystyle \frac {K^{2}\,(6 - K^{2})}{{k_{1}}\,{k_{2}}}}  &
 - {\displaystyle \frac {1}{3}} \,{\displaystyle \frac {K^{2}\,(6
 - K^{2})}{{k_{2}}^{2}}} 
\end{array}
 \right] 
\end{eqnarray*}
where $\bar K \equiv \sqrt{K^{2}\,(6 - K^{2})}$.

The characteristic polynomial of the matrix $D{\bf F}(A)$ can be simplified to the form
$$
c_A(\lambda)= \lambda ^{4} 
+ (8 - 2\,K^{2})\lambda ^{3} 
+ (\frac {3}{4}K^{4} + 21 - 8K^{2})\lambda ^{2} 
+ (\frac {1}{2}K^{6} - \frac {7}{2}K^{4} + 18)\lambda  
- \frac {1}{4}K^{8} + 18K^{2} - 15K^{4} + \frac {7}{2}K^{6}
$$
The roots of $c_A(\lambda)$ yield the eigenvalues seen in equation (3.5).  The eigenvalues of all the other equilibrium points can be computed in a similar fashion as above.


\squeezetable
\begin{table}
\caption{Equilibrium points  in the two-scalar field model with no
matter.  In the table, $A$ and $CS$ correspond to the {\it
two-field} assisted inflationary solution and the {\it two-field}
curvature scaling, solution respectively.} \label{table1}
\begin{tabular}{@{}llccl}
Solution/Label  & Coordinates       & Deceleration      & Curvature & Eigenvalues \\
\ & $\{\Psi_1, \Psi_2, \Phi_1, \Phi_2\}$ & Parameter, $q$   & $^3R/H^2$  \\
\hline
    Assisted Inflation, $A$ &
    $\{\Psi_1, \Psi_2, \Phi_1, \Phi_2\}^A$ (see \ref{AAA})&
    $q_A$ (see \ref{qA})&
    $0$ &
    (see \ref{eigen})\\
\ \\
    Power-Law, $P_1$ &
    $\displaystyle \{-\frac{k_1}{\sqrt{6}}, 0 , \sqrt{1-\frac{k_1^{\ 2}}{6}}, 0 \} $ &
    $\displaystyle \frac{k_1^{\ 2}-2}{2}$ &
    $0$ &
    $\displaystyle k_1^{\ 2}-2,\frac{k_1^{\ 2}}{2},\frac{k_1^{\ 2}-6}{2},\frac{k_1^{\ 2}-6}{2}$ \\
\ \\
    Power-Law, $P_2$ &
    $\displaystyle \{0, -\frac{k_2}{\sqrt{6}}, 0 , \sqrt{1-\frac{k_2^{\ 2}}{6}} \} $ &
    $\displaystyle \frac{k_2^{\ 2}-2}{2}$ &
    $0$ &
    $\displaystyle k_2^{\ 2}-2,\frac{k_2^{\ 2}}{2},\frac{k_2^{\ 2}-6}{2},\frac{k_2^{\ 2}-6}{2}$ \\
\ \\
    Curvature Scaling, $CS$ &
    $\displaystyle \{-\frac{\sqrt{6}}{3k_1},-\frac{\sqrt{6}}{3k_2},
    \frac{2}{\sqrt{3}k_1} , \frac{2}{\sqrt{3}k_2} \} $ &
    $0$ &
    $\displaystyle \frac{2(k_1^{\ 2}+k_2^{\ 2})-k_1^{\ 2}k_2^{\ 2}}{k_1^{\ 2}k_2^{\ 2}}$ &
    $\displaystyle -1\pm\sqrt{3}i,$ \\
    \ & \ & \ & \ & $\displaystyle -1 \pm \sqrt{1+4[2(k_1^{-2}+k_2^{-2})-1])}$ \\

\ \\
    Curvature Scaling, $CS_1$ &
    $\displaystyle \{-\frac{\sqrt{6}}{3k_1}, 0 , \frac{2}{\sqrt{3}k_1},0 \} $ &
    $0$ &
    $\displaystyle \frac{2-k_1^{\ 2}}{k_1^{\ 2}}$ &
    $\displaystyle -2, 1, -1 \pm \sqrt{1+4k_1^{-2}(2-k_1^{\ 2})}$ \\
\ \\
    Curvature Scaling, $CS_2$ &
    $\displaystyle \{0,-\frac{\sqrt{6}}{3k_2}, 0 , \frac{2}{\sqrt{3}k_2} \} $ &
    $0$ &
    $\displaystyle \frac{2-k_2^{\ 2}}{k_2^{\ 2}}$ &
    $\displaystyle -2, 1, -1 \pm \sqrt{1+4k_2^{-2}(2-k_2^{\ 2})}$ \\
\ \\
    Milne, $M$ &
    $\displaystyle \{0, 0, 0, 0\}$ &
    $\displaystyle 0$ &
    $-1$ &
    $-2,-2,1,1$ \\
\ \\
    Massless Scalar Field, $MSF$ &
    $\displaystyle \{\Psi_0, \epsilon\sqrt{1-\Psi_0^{\ 2}},0,0\} $ &
    $2$ &
    $0$ &
    $\displaystyle 0,4, 3+\frac{\sqrt{6}}{2}k_1\Psi_0, 3+\frac{\sqrt{6}}{2}\epsilon k_2 \sqrt{1-\Psi_0^{\ 2}}$ \\
    \ & where $0\leq\Psi_0^{\ 2}\leq 1$ & & & where $\epsilon = \pm 1$.\\
\ \\
\end{tabular}
\end{table}

\newpage

\squeezetable
\begin{table}
\caption{Equilibrium points with $\Omega=0$ in the two-scalar field model with matter.
Each equilibrium point has $\Omega=0$ and the coordinates given in Table \ref{table1}.
The additional fifth eigenvalue is displayed.} \label{table2}
\begin{tabular}{@{}lcccccccc}
Label  & $A$ & $P_1$ & $P_2$ & $CS$ & $CS_1$ & $CS_2$ & $M$ &
$MSF$ \\
\ \\
5$^{th}$ Eigenvalue     & $\frac{k_1^{\ 2}k_2^{\
2}}{k_1^{\ 2}+k_2^{\ 2}} -3\gamma $
            & $k_1^{\ 2}-3\gamma$
            & $k_2^{\ 2}-3\gamma$
            & $2-3\gamma$
            & $2-3\gamma$
            & $2-3\gamma$
            & $2-3\gamma$
            & $3(2-\gamma)$
\end{tabular}
\end{table}

\vglue 1.5in

\squeezetable
\begin{table}
\caption{Equilibrium points with $\Omega \ne 0$ in the two-scalar field model with matter.
 Note that in each case $^3R=0$ and $q=(3\gamma-2)/2$.} \label{table3}
%
%
\begin{tabular}{@{}lll}
Solution/Label  & Coordinates            & Eigenvalues \\
\ & $\{\Omega,\Psi_1, \Psi_2, \Phi_1, \Phi_2\}$     &   \\
\hline
    FRW, $F$ &
    $\displaystyle \{1, 0, 0, 0, 0\}$ &
    $3\gamma-2, \frac{3}{2}(\gamma-2),
    \frac{3}{2}(\gamma-2),\frac{3}{2}\gamma,\frac{3}{2}\gamma$ \\
\ \\
    Matter Scaling, $MS_1$ &
    $\displaystyle \{1-\frac{3\gamma}{k_1^{\ 2}},-\frac{\sqrt{6}\gamma}{2k_1}, 0 ,
        \frac{\sqrt{6\gamma(2-\gamma)}}{2k_1}, 0 \} $ &
    $\displaystyle  \frac{3}{2}(\gamma-2), \frac{3}{2}\gamma, 3\gamma-2,$\\
    & & $\displaystyle
    \frac{3}{4}\left( (\gamma-2)\pm \sqrt{(\gamma-2)^2+8\gamma(\gamma-2)[1-3\gamma k_1^{\ -2}]}\right)$ \\
\ \\
    Matter Scaling, $MS_2$ &
    $\displaystyle \{1-\frac{3\gamma}{k_2^{\ 2}},0,-\frac{\sqrt{6}\gamma}{2k_2}, 0 ,
        \frac{\sqrt{6\gamma(2-\gamma)}}{2k_2}\} $ &
    $\displaystyle  \frac{3}{2}(\gamma-2), \frac{3}{2}\gamma, 3\gamma-2,$\\
    & & $\displaystyle
    \frac{3}{4}\left( (\gamma-2)\pm \sqrt{(\gamma-2)^2+8\gamma(\gamma-2)[1-3\gamma k_2^{\ -2}]}\right)$ \\
\ \\
    Matter Scaling, $MS$ &
    $\displaystyle \{1-3\gamma\left(k_1^{\ -2}+k_2^{\ -2}\right),
    -\frac{\sqrt{6}\gamma}{2k_1}, -\frac{\sqrt{6}\gamma}{2k_2}$ ,&
    $\displaystyle 3\gamma-2,      
    \frac{3}{4}\left( (\gamma-2)\pm \sqrt{(\gamma-2)^2+8\gamma(\gamma-2)}\right)$\\
    &
    $ \displaystyle \frac{\sqrt{6\gamma(2-\gamma)}}{2k_1}, \frac{\sqrt{6\gamma(2-\gamma)}}{2k_1} \} $ &
    $\displaystyle
    \frac{3}{4}\left( (\gamma-2)\pm \sqrt{(\gamma-2)^2+8\gamma(\gamma-2)[1-3\gamma( k_1^{\ -2}+ k_2^{\ -2})]}\right)$ \\
\end{tabular}
\end{table}


\begin{references}


\bibitem{Guth} A. A. Starobinsky, Phys. Lett. B {\bf 91}, 99 (1980);
A. H. Guth, Phys. Rev. D {\bf 23}, 347 (1981); A. D. Linde, Phys. Lett.
B {\bf 108}, 389 (1982).

\bibitem{Olive90} K.A. Olive, Phys. Rep. {\bf 190}, 308
(1990).

\bibitem{Luchin85} F. Lucchin and S. Matarrese, Phys. Rev. D
{\bf 32}, 1316 (1985).

\bibitem{Wetterich} C. Wetterich, Nucl. Phys. {\bf B302}, 668 (1988).

\bibitem{BC99} A.P. Billyard and A.A. Coley, Phys. Rev. D {\bf x}, x (1999).

\bibitem{Halliwell87} J.J. Halliwell, Phys. Lett. B {\bf
185}, 341 (1987);
A.B. Burd and J.D. Barrow, Nucl. Phys. B
{\bf 929} (1988);
Y. Kitada and K. Maeda, Class. Quantum Grav. {\bf 10},  703
(1993).

\bibitem{Coley97a}
A.~A. Coley, J. Ib{\'a}{\~n}ez, and R.~J. van~den Hoogen, J.
Math. Phys. {\bf
  38},  5256  (1997);
R.~J. van~den Hoogen, I. Olasagasti, Phys. Rev. D {\bf
  59},  107302  (1999).

\bibitem{Billyard99} A.P. Billyard, A.A. Coley, R.J. van den
Hoogen, J. Ib\'a\~nez, and I. Olasagasti, accepted for
publication in Class. Quantum Grav. (1999).

\bibitem{Copeland98} E.J. Copeland, A.R. Liddle, and D.
Wands, Phys. Rev. D {\bf 57}, 4686 (1998).

\bibitem{Billyard98} A.P. Billyard, A.A. Coley, and R.J. van
den Hoogen, Phys. Rev. D {\bf 58}, 123501 (1998).

\bibitem{vandenHoogen99} R.J. van den Hoogen, A.A. Coley,
and D. Wands, Class. Quant. Grav. {\bf 16}, 1 (1999).

\bibitem{Green} M. B. Green, J. H. Schwarz, and E. Witten, {\em Superstring
Theory} (Cambridge University Press, 1987).

\bibitem{Liddle98} A.R. Liddle, A. Mazumdar and F.E.
Schunck, Phys. Rev. D {\bf 58}, 061301, (1998) [astro-ph/9804177].

\bibitem{Malik98} K. Malik and D. Wands, Phys. Rev. D {\bf 59}, 123501 (1999)
[astro-ph/9812204].

\bibitem{hybrid1} A. D. Linde, Phys. Lett. {\bf B259}, 38 (1991);

\bibitem{vandenHoogen99d}
R.~J. van~den Hoogen and L. Filion, submitted to Class. Quantum Grav., Preprint  (1999).

\bibitem{Berkin90} A. L. Berkin, K. Maeda, and J. Yokoyama,
Phys. Rev. Lett. {\bf 65}, 141 (1990);
G. Abolghasem, A. Burd, A. Coley, R.
van den Hoogen, Phys. Rev. D {\bf 48}, 557 (1993).

\bibitem{Copeland99} E.J. Copeland, A. Mazumdar, and N.J.
Nunes, [astro-ph/9904309].

\bibitem{stelle} M. Bremer, M.J. Duff, H. Lu, C.N. Pope, K.S. Stelle,
Nucl.Phys. B {\bf 543} 321 (1999).

\bibitem{reall} S.W. Hawking and H. Reall,
Phys. Rev D {\bf 59}, 023502 (1999).

\bibitem{GL} A. M. Green and J. E. Lidsey, [astro/99007223].

\bibitem{ko}  P. Kanti and K. A. Olive, Phys. Rev. D {\bf 60}, 043502 (1999)
[hep-ph/9906331].

\bibitem{Kanti99} P. Kanti and K.A. Olive, preprint,
hep-ph/9903524 (1999).


\bibitem{KL} N. Kaloper and A. R. Liddle, [hep-ph/9910499]

\bibitem{WainwrightEllis97}  J. Wainwright and G.F.R. Ellis,
{\em Dynamical Systems in Cosmology}, (Cambridge University
Press, 1997).

\bibitem{LuPope}  H. L\"u and C. N. Pope, Nucl. Phys. {\bf B465}, 127 (1996).

\bibitem {doubleinflate} L. A. Kofman, A. D. Linde and A. A. Starobinsky, Phys. Letts. B, {\bf 157}, 361 (1985); K. Yamamoto, M. Sasaki, and T. Tanaka, Astrophysical Journal, {\bf 455}, 412 (1995).

\bibitem{FJ} P.G. Ferreira and M. Joyce, Phys. Rev. Lett. {\bf 79},
4740 (1997) and Phys. Rev. D{\bf58}, 023503 (1998); C. Wetterich,
Astron. Astrophys. {\bf 301}, 321 (1995).

\bibitem{vandenHoogen97a}
R.~J. van~den Hoogen, A.~A. Coley, and J. Ib{\'a}{\~n}ez, Phys. Rev. D {\bf
  55},  1  (1997).

\bibitem{Feinstein93} A. Feinstein and J. Ib\'{a}\~{n}ez,  Class. Quantum Grav.  {\bf 10}, 93 (1993).

\bibitem{DS} S. Wiggins, {\em Introduction to Applied Nonlinear Dynamical Systems and Chaos}, (Springer-Verlag, New York, 1990);  J.H. Hubbard and B.H. West, {\em Differential Equations: A Dynamical Systems Approach}, (Springer-Verlag, New York, 1991); L. Perko, {\em Differential Equations and Dynamical Systems}, (Springer-Verlag, New York, 1996).

\end{references}
\end{document}